\documentclass[12pt]{article}

\usepackage{amsmath}
\usepackage{amssymb}
\usepackage{amsfonts}
\usepackage{graphicx}

\numberwithin{equation}{section}

\begin{document}

\begin{center}
{\large \bf{ The Higgs condensate as a quantum liquid }}
\end{center}

\vspace*{2cm}

\begin{center}
{
Paolo Cea~\protect\footnote{Electronic address:
{\tt paolo.cea@ba.infn.it}}  \\[0.5cm]
{\em INFN - Sezione di Bari, Via Amendola 173 - 70126 Bari,
Italy} }
\end{center}

\vspace*{1.5cm}

\begin{abstract}
\noindent 
 We model the Higgs condensate of the Standard Model as a relativistic quantum fluid analogous
 to superfluid helium. We find that the low-lying excitations
 of the Higgs condensate behave like two relativistic Higgs fields. The lighter Higgs boson has a mass of order
 $10^2$ GeV.  We identify this light Higgs particle with the new LHC resonance at 125 GeV. The heavy Higgs boson
 has a mass around 750 GeV consistent with our recent phenomenological analysis  of the preliminary LHC Run 2 data
in the golden channel. We critical compare our theoretical scenario with two Higgs bosons to the available LHC Run 2 data.
\vspace{0.5cm}

\noindent
 {\it Keywords}: Higgs Boson; Large Hadron Collider
 \vspace{0.2cm}

\noindent
 PACS Nos.: 11.15.Ex; 14.80.Bn; 12.15.-y

\end{abstract}
\newpage
\noindent
\section{Introduction}
\label{s-1}
It is known since long time that in the Standard Model, within the non-perturbative description of spontaneous 
symmetry breaking~\cite{Englert:1964,Higgs:1964,Guralnik:1964,Higgs:1966},
self-interacting scalar fields suffer the triviality problem~\cite{Fernandez:1992}, namely the renormalised self-coupling
goes to zero when the ultraviolet cutoff  is sent to infinity.
Nevertheless, extensive numerical simulations showed that, even without self-interactions, the scalar bosons could
trigger  spontaneous  symmetry breaking. Moreover, precise non-perturbative numerical simulations~\cite{Cea:2004,Cea:2012}
indicated that the excitation of the Bose-Einstein scalar condensate is a rather heavy scalar particle. In fact, our recent analysis
of the preliminary LHC Run 2 data in the so-called golden channel~\cite{Cea:2017,Cea:2019} (see also Ref.~\cite{Richard:2020}),
 showed a rather convincing evidence of  a  broad scalar resonance with mass around 730 GeV, that  seems to be consistent 
 with a heavy Higgs boson. \\
Supposing that the full Run 2 data set will confirm the heavy Higgs boson proposal, we face with the problem of the existence of two Higgs bosons
considering that  the first runs of proton-proton collisions at the CERN Large Hadron Collider  with center-of-mass
energies $\sqrt{s} = 7 $  and  $8 \;$ TeV (Run 1) gave evidence  for 
a spin-zero boson  with mass 125  GeV~\cite{Aad:2012,Chatrchyan:2012}, and that it is now well 
established that this narrow resonance resembles closely the 
Higgs boson of the Standard Model~\cite{Aad:2019,CMS:2019}.   \\
In the present paper we propose to look at the Higgs condensate as a quantum liquid analogous to the Bose-Einstein condensate
in superfluid helium (helium II)~\footnote{ For a good account, see  Refs.~\cite{Keller:1969,Galasiewicz:1971}.}. We find that  the low-lying excitations of the Higgs condensate resemble two Higgs bosons with masses
of order 100 GeV and around 750 GeV, respectively. These condensate excitations parallel the phonons and rotons in superfluid helium. \\
The remainder of the paper  is organised as follows. In Sect.~\ref{s-2} we briefly review the main properties of liquid helium in the
superfluid phase. In Sect.~\ref{s-3} we discuss the Higgs mechanism taking into account the problem of triviality for self-interacting
scalar fields in (3+1)-dimensions. The presence of two Higgs bosons is addressed in Sect.~\ref{s-4}. 
Sect.~\ref{s-5} is devoted to the phenomenological signatures  of the two  Higgs bosons and to a critical comparison with available
experimental observations. Finally, in Sect.~\ref{s-6} we summarise the main results of the paper.
\section{The helium II }
\label{s-2}
The condensation of a relativistic scalar field free asymptotically could appear paradoxical. Notwithstanding, in condensed matter physics it is
known that an ideal non-relativistic Bose gas does display the Bose-Einstein condensation at sufficiently low temperatures. Indeed,
in a non-interacting boson system at absolute zero temperature all particles will be in the state of zero momentum. An excitation of momentum
$\vec{p}$ will possess  the free-particle energy $\varepsilon_{\vec{p}} = \vec{p}^{~2}/2 m$, $m$ being the particle mass.
 However, note that,  as we shall see later on, when the interactions between bosons are taken into account the quasi-particle excitation
 spectrum is drastically altered. 
 \\
 Soon after the remarkable discovery of superfluidity in liquid helium below the so-called
 $\lambda$-point~\cite{Keller:1969,Galasiewicz:1971},  it was suggested that helium II should be considered as a degenerate
 ideal Bose gas that, indeed, manifests the Bose-Einstein condensation at a temperature close to the observed critical
 temperature~\cite{London:1938a,London:1938b,Tisza:1938,London:1954}. However, Landau~\cite{Landau:1941,Landau:1947} pointed out that the
  suggestion by London and Tisza cannot account for the superfluidity of helium II below the $\lambda$-point.
  In fact,  the remarkable properties of superfluid helium could be recovered if helium II were composed of an intimate mixture of two
  fluid, one fluid with zero viscosity and the other with normal viscosity.  Landau~\cite{Landau:1941,Landau:1947} developed a peculiar two-fluid
 hydrodynamics model in which  he explained the phenomenon of superfluidity as a consequence of an excitation spectrum of helium II
 derived empirically.  Remarkably, the Landau two-fluid theory and the empirically derived excitation curve explain a great many
 of the superfluid properties of liquid helium. Actually, Landau assumed that every weakly excited state of helium II could be
 considered as an aggregate of elementary excitations. The potential motion of the quantum fluid was  assumed to be
 due to sound waves and the corresponding elementary excitations were the phonons with a linear dispersion form:
\begin{equation}
\label{2.1}
\varepsilon_{ph}(\vec{p})  \; =  \;  c_s \;  | \vec{p} |  \; , 
\end{equation}
where $c_s$ is the sound velocity.  The vortex motion of the fluid was ascribed to gapped elementary excitations, called rotons,  
with dispersion law:
\begin{equation}
\label{2.2}
\varepsilon_{rot}(\vec{p})  \; =  \;  \Delta \; + \frac{(\vec{p}^{\, 2} - \vec{p}_0^{\, 2})}{2 \, m^*} \; ,
\end{equation}
where $\Delta$ is a constant, $m^*$ some effective mass, and $\vec{p}_0$ is a momentum of order $|\vec{p}_0| \, \sim \, 1/d$ with
$d$ the average distance between helium atoms, i.e. $d \, \simeq \, n^{-1/3}$, $n$ being the number density. \\
Bogoliubov~\cite{Bogoliubov:1947} attempted to explain the phenomenon of superfluidity on the basis of the theory of Bose-Einstein
condensation in a non-perfect gas. In fact, Bogoliubov considered a Bose gas with short-range repulsive interactions
characterised by the s-wave scattering length $a_s$ in the dilute gas approximation, $a_s \, n^{1/3} \; \ll \; 1$.
Using the methods of second quantisation and a new perturbative technique, Bogoliubov was able to show that the existence and the properties
of the elementary excitations followed directly from the quantum-mechanical equations describing the Bose-Einstein condensation
of the non-ideal gas. Moreover, Bogoliubov showed that the low excited states of the Bose gas can be described as a perfect 
Bose-Einstein gas of phonons. Finally, Feynman~\cite{Feynman:1954,Feynman:1955,Feynman:1956} showed that the excitation
spectrum postulated by Landau could be derived within a first principle quantum-mechanical approach. Actually, Feynman
convincingly showed that the only low-energy excitations in helium II were the Bogoliubov's phonons. In addition,  it turned out 
that from the microscopic point of view a roton may be considered like a small vortex ring. Therefore, rotons corresponds to high-energy
 excitations of the condensate localised on a region of order of the average distance between helium atoms. \\
 To summarise,  the remarkable superfluid behaviour of the helium II quantum liquid can be understood if the excitation spectrum
 is an almost ideal gas of elementary quasi-particles. The low-energy elementary excitations are the Bogoliubov's phonons that
 are collective excitations that retain the needed quantum coherence for wavelengths $\lambda_{ph} \, \gg \, d$. The high-energy excitations
 are rotons, namely localised excitations of the Bose-Einstein condensate with wavelength of the order of the 
 distance between helium atoms.
\section{The scalar condensate}
\label{s-3}
The Higgs mechanism in the Standard Model is implemented by the Bose-Einstein condensation of a relativistic scalar field. To illustrate
in the simplest way the mechanism let us consider a real scalar field defined by the Lagrangian density:
\begin{equation}
\label{3.1}
 {\cal L}(x) \; = \;  \frac{1}{2} \; \partial_{\mu}  \phi(x) \,   \partial^{\mu}  \phi(x) \;  - \;   \frac{1}{2} \; m^2  \phi^2(x) \;
  -  \;  \frac{\lambda}{4!} \, \phi^4(x)  \; ,
\end{equation}
with $\lambda > 0$. For $m^2 > 0$, the Lagrangian Eq.~(\ref{3.1}) describes a self-interacting scalar field with bare mass $m$. To implement
the Bose-Einstein condensation we must assume $m^2 < 0$. In this case there is a macroscopic occupation of the zero mode of the
scalar field. Accordingly, the vacuum expectation value of the quantum field $\hat{\phi}(x)$ is different from zero:
\begin{equation}
\label{3.2}
<0| \; \hat{\phi} \; |0> \; =  \; v \; .
\end{equation}
Therefore we are led to write:
\begin{equation}
\label{3.3}
\phi(x) \;  =  \; h(x) \; + \; v \; ,
\end{equation}
so that  $<v|\hat{h} |v>  = 0$. Rewriting the Lagrangian Eq.~(\ref{3.1}) in terms of the shifted field $h(x)$ we get:
\begin{equation}
\label{3.4}
 {\cal L}(x) \; = \;  \frac{1}{2} \; \partial_{\mu}  h(x) \,   \partial^{\mu}  h(x) \;  - \;   \frac{1}{2} \; m_h^2 \, h^2(x) \;
 -  \;  \frac{\lambda \, v}{6} \, h^3(x)  \;  -  \;  \frac{\lambda}{4!} \, h^4(x)  \; ,
\end{equation}
with:
\begin{equation}
\label{3.5}
 m_h^2 \; = \; \frac{1}{3} \, \lambda \, v^2 \; ,
\end{equation}
while the correct vacuum expectation value $v$ corresponds to the vanishing of the tadpole. This perturbative implementation
of the Bose-Einstein condensation of a relativistic scalar field is the generally accepted procedure in high energy physics.
The standard perturbative scheme parallels closely the Bogoliubov's perturbative approximation. The elementary excitations of the
scalar condensate are given by a quantum scalar field $\hat{h}(x)$ with mass given by Eq.~(\ref{3.5}) and cubic and quartic self-couplings.
Thus, these elementary excitations are coherent long-range collective excitations of the scalar condensate that are analogous to
the phonons in helium II. In the following we shall call these excitations the Bogoliubov's branch of the condensate excitation spectrum.
It is worthwhile to observe that the presence of the Bogoliubov's branch is assured by the short-range repulsive interaction given by
the positive quartic self-interaction term. Within the Bogoliubov's perturbative approximation there is no way to recover the roton branch
of the excitation spectrum. However, this perturbative scheme is doomed to failure since self-interacting scalar fields are subject to
the triviality problem~\cite{Fernandez:1992}, i.e. the renormalised self-coupling $\lambda \rightarrow 0$ when the ultraviolet cutoff
is sent to infinity. If this is the case, the Lagrangian Eq.~(\ref{3.1}) should reduce to the Lagrangian of a free scalar field.
Naively one expects that the spontaneous symmetry breaking mechanism cannot be implemented without the scalar quartic self-coupling.
However, one should keep in mind that a non-relativistic ideal Bose gas does develop the Bose-Einstein condensation. In the case of a trivial
relativistic scalar field the onset of the condensation phase is given by the vanishing of the mass term $m^2=0$.  Writing:
\begin{equation}
\label{3.6}
\phi(x) \;  =  \; H(x) \; + \; v \; ,
\end{equation}
extensive numerical studies~\cite{Cea:2004,Cea:2012} showed that the fluctuating field $H(x)$ behaves as a free massive scalar field
with mass finitely related to $v$:
\begin{equation}
\label{3.7}
m_H  \;  =  \; \xi  \; v \; .
\end{equation}
Moreover, it turned out that in the continuum limit~\cite{Cea:2004,Cea:2012} :
\begin{equation}
\label{3.8}
  \xi  \; =  \;  3.07 \; \pm \; 0.11 \; ,
\end{equation}
where the uncertainties include both the statistical and systematic errors. Assuming that $v$ is the known weak scale of the Standard Model:
\begin{equation}
\label{3.9}
 v \; \simeq \; 246  \;  {\text GeV} \; , 
\end{equation}
from Eqs.~(\ref{3.8}) and (\ref{3.9}) we get:
\begin{equation}
\label{3.10}
m_H  \;  =  \; 756  \; \pm \; 28 \;  {\text GeV} \; .
\end{equation}
We see, then, that the excitations over the condensate are like the rotons in superfluid helium. The Bogoliubov's branch of the excitation
spectrum is absent since the triviality of the theory implies $\lambda = 0$.
\section{Two  Higgs bosons}
\label{s-4}
In the previous Section we have discussed the Bose-Einstein condensation for a relativistic real scalar field. Obviously, one could object that
our discussion is not directly related  to the Standard Model since the relevant  scalar sector is the $O(4)$-symmetric self-interacting
scalar theory.  However, the known Higgs mechanism eliminates three scalar fields (the Goldstone bosons) leaving as the physical
Higgs field the radial excitation whose dynamics is described by the one-component (i.e. real) self-interacting scalar field theory. \\
We said that a real self-interacting scalar field is trivial, namely it is a free field asymptotically when the ultraviolet cutoff is sent to infinity.
Even though a rigorous proof of triviality in (3+1)-dimensions is lacking, there are several convincing numerical studies that leave little
doubt on the triviality of the scalar theories. {\it Rebus sic standibus}, the elementary excitations of the Higgs condensate should be a massive
scalar field with a rather heavy mass given by Eq.~(\ref{3.10}). In our previous paper~\cite{Cea:2019} we identified this elementary
excitation as the true Higgs mode. As a matter of fact, we showed~\cite{Cea:2019} that there is some evidence of this Higgs mode in
the so-called golden channel. Nevertheless, it is widely accepted that the Higgs boson is the new narrow resonance at 125 GeV detected
by both the ATLAS and CMS Collaborations~\cite{Aad:2012,Chatrchyan:2012}. In the present Section we will show that there are
two Higgs bosons. The true Higgs mode is the heavy resonance already discussed, while the resonance at 125 GeV is the light Higgs boson
that correspond to the Bogoliubov's branch of the excitation spectrum of the Higgs condensate.  To see  this we need a non-zero quartic 
self-interaction of the Higgs field. \\
Due to the triviality of self-interacting scalar fields the Higgs mode can interact only through the couplings to gauge and fermion fields.
In fact, the interactions with vector bosons and fermion fields will induce an effective scalar self-coupling. If we define the effective
quartic term in the Higgs potential:
\begin{equation}
\label{4.1}
V^{(4)}  \;  = \; \frac{\lambda_{eff}}{6}  \; \left [ \Phi^{\dagger}(x) \, \Phi(x) \right ]^2  \; \; , \; \; \Phi(x) \; = \;   
 \frac{1}{\sqrt{2}} \; \begin{pmatrix} 0 \\ H(x) \; + \; v \end{pmatrix}  \;  \; ,
\end{equation}
then the renormalisation-group equation for the self-coupling $\lambda_{eff}$ can be easily obtained following Refs.~\cite{Cheng:1975,Cabibbo:1979}.
\begin{figure}
\includegraphics[width=1.0\textwidth,clip]{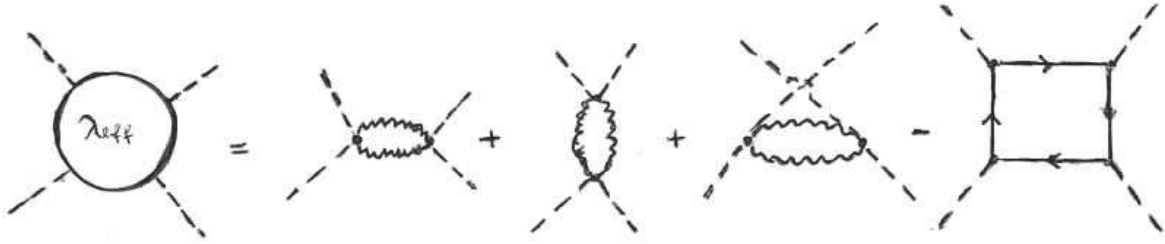}
\caption{\label{Fig1} 
Lowest-order contributions to the quartic  self-coupling renormalisation.}
\end{figure}
In the lowest approximation the relevant Feynman diagrams are displayed in Fig.~\ref{Fig1}. We have:
\begin{equation}
\label{4.2}
 \frac{ d \lambda_{eff}(t)}{d t }  \; \simeq \; \frac{1}{16 \pi^2} 
 \left \{ \frac{9}{4} \left [ 2 g^4 + (g^2+g'^2)^2 \right ]  - 9 \, \kappa^2 \lambda_t^4 \right \}  \; . 
\end{equation}
In Eq.~(\ref{4.2}) $g$ and $g'$ are the couplings to the weak $SU(2)$ and $U(1)$ respectively, which are related to the
electric charge according to the well-known formula:
\begin{equation}
\label{4.3}
g \; = \;   \frac{e}{sin \, \theta_W} \; \; , \; \; g' \; = \; g \; tg \, \theta_W \; = \; \frac{e}{cos \, \theta_W} \; ,
\end{equation}
$\theta_W$ being the Weinberg's angle. According to our previous paper~\cite{Cea:2019}, we are considering only the Yukawa
coupling to the top quark, $\lambda_t = \frac{\sqrt{2} \, m_t}{v}$,  where $m_t \simeq 173$ GeV is the top mass. 
Moreover, our phenomenological analysis suggested that the coupling of the Higgs mode to the top quark were strongly
suppressed such that:
\begin{equation}
\label{4.4}
\lambda_t^2 \; \rightarrow \; \kappa \, \lambda_t^2 \; \; , \; \; \kappa \; \simeq \; 0.15 \; \; .
\end{equation}
Finally, in Eq.~(\ref{4.2}) we set $t = ln (M/\mu)$, where $\mu \ll v$. \\
By solving   Eq.~(\ref{4.2}) one gets the effective self-coupling at the scale $M > \mu$ once the couplings are fixed at the starting scale $\mu$.
We note that the triviality of the Higgs scalar field assures that $\lambda_{eff}(\mu) \simeq 0$. Therefore, to the lowest-order approximation
we obtain:
\begin{equation}
\label{4.5}
\lambda_{eff}(M)  \; \simeq \; \frac{1}{16 \pi^2} 
 \left \{ \frac{9}{4} \left [ 2 g^4(\mu) + (g^2(\mu)+g'^2(\mu))^2 \right ]  - 9 \, \kappa^2 \lambda_t^4(\mu) \right \}  ln (M/\mu) \; . 
\end{equation}
It is useful to rewrite this last equation as:
\begin{eqnarray}
\label{4.6}
\lambda_{eff}(M)  \; \simeq \; 
\left  \{ \frac{9}{4} \,  \alpha_{QED}(\mu) \left [ \frac{2}{sin^4 \, \theta_W(\mu)} + 
  \left ( \frac{1}{sin^2 \, \theta_W(\mu)} + \frac{1}{cos^2 \, \theta_W(\mu)} \right )^2 \right ]  \right .
  \nonumber \\
\left .  -  \; \frac{9}{16 \pi^2} \, \kappa^2 \lambda_t^4(\mu) \right \}  ln (M/\mu) \; . 
\end{eqnarray}
Once we have an effective self-coupling, within the Bogoliubov's  approximation, we recover the phonon branch of the condensate
excitation spectrum that behaves like a scalar field $h(x)$. Since the Bose-Einstein condensation sets in at $m^2=0$, the mass
of the scalar field $h(x)$ is now:
\begin{equation}
\label{4.7}
m^2_h   \; =  \; \frac{1}{2} \; \lambda_{eff}(M)  \; v^2 \; .  
\end{equation}
To avoid confusion or misunderstanding,  it is necessary to pause and add some comments on our results.  We are not saying that
there are two different elementary Higgs fields. On the contrary, we have a unique quantum Higgs field. However, since the scalar
condensate behaves like the helium II quantum liquid, when the Higgs field acts on the condensate it can give rise to two elementary excitations,
namely the phonon-like and roton-like excitations corresponding to long-range collective and localised disturbances of the condensate, respectively.
These elementary condensate excitations behave as weakly interacting scalar fields with vastly different mass.  The main advance in our approach
is that the Higgs boson masses are not  free parameters, but these can be estimated from first principles. \\
To complete the mass calculations we must consider the effects due to the vector bosons and fermions. In the one-loop approximation
we have the mass corrections displayed in Fig.~\ref{Fig2}. 
\begin{figure}
\includegraphics[width=1.0\textwidth,clip]{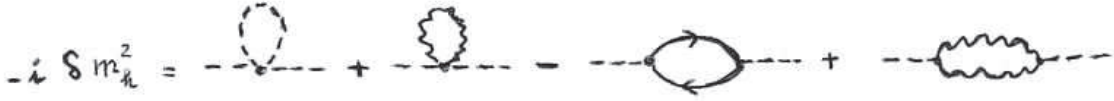}
\caption{\label{Fig2} 
Lowest-order contributions to the scalar field mass term. }
\end{figure}
Note that in Fig.~\ref{Fig2} we are neglecting the one-loop term of order $\lambda_{eff}^2$, since it is easy to check that $\lambda_{eff} \ll 1$.
In principle, the one-loop diagrams in Fig.~\ref{Fig2} should modify the mass of the elementary excitations. Within the approximation of weakly 
interacting condensate excitations we may consider the phonon-like and roton-like excitations as independent scalar particles.
Now, we have seen that the Bogoliubov's  branch is composed by long-range collective excitations of the Higgs condensate that, therefore,
retain the needed coherence to propagate for wavelengths up to the size of the roton-like excitations $\sim 1/m_H$.
So that, in evaluating the mass corrections we have that the integrals with loop momenta $k \lesssim  \Lambda_h \simeq m_H$
are to be considered as finite mass corrections to the $h(x)$ scalar field. On the other hand, for loop momenta $k  >  \Lambda_h$
the resulting mass corrections (that contain quadratic and logarithmic divergences) must be incorporated into the mass term of the
$H(x)$ field. The only effect is that one must tune the bare mass term until  the total mass is set to zero to ensure the onset  of
the Bose-Einstein condensation. As a consequence, we are left with:
\begin{equation}
\label{4.8}
m^2_h   \; =  \; \frac{1}{2} \; \lambda_{eff}  \; v^2 \;  + \; \delta  m^2_h    \; ,
\end{equation}
where:
\begin{equation}
\label{4.9}
 \delta  m^2_h \; = \;   \delta  m^2_W \; + \;  \delta  m^2_Z \; + \;   \delta  m^2_{\lambda_{eff}} \; + \;    \delta  m^2_t  \; . 
\end{equation}
We find:
\begin{eqnarray}
\label{4.10}
 \delta  m^2_W  \;  \simeq  \; 
\frac{3}{16 \pi} \,  \frac{\alpha_{QED}(\Lambda_h)}{sin^2 \, \theta_W}  \left [  \Lambda_h^2 - M_W^2 \;  ln \left(  \frac{\Lambda_h^2+M_W^2}{M^2_W } \right ) \right ] \;  + \;
   \nonumber \\
 \frac{3}{8 \pi} \, \frac{\alpha_{QED}(\Lambda_h)}{sin^2 \, \theta_W} \;  M_W^2  \left [  1 -  \;  ln \left(  \frac{\Lambda_h^2+M_W^2}{M^2_W } \right ) 
 \; - \;  \frac{M^2_W }{\Lambda_h^2+M_W^2}  \right ]  \; ,
\end{eqnarray}
\begin{eqnarray}
\label{4.11}
 \delta  m^2_Z  \;  \simeq  \; 
\frac{3}{32 \pi} \,  \frac{\alpha_{QED}(\Lambda_h)}{sin^2 \, \theta_W \, cos^2 \, \theta_W} 
 \left [  \Lambda_h^2 - M_Z^2 \;  ln \left(  \frac{\Lambda_h^2+M_Z^2}{M^2_Z } \right ) \right ] \;  + \;
   \nonumber \\
 \frac{3}{8 \pi} \, \frac{\alpha_{QED}(\Lambda_h)}{sin^2 \, \theta_W  \, cos^2 \, \theta_W} \;  M_Z^2 
  \left [  1 -  \;  ln \left(  \frac{\Lambda_h^2+M_Z^2}{M^2_Z } \right ) 
 \; - \;  \frac{M^2_Z }{\Lambda_h^2+M_Z^2}  \right ]  \; ,
\end{eqnarray}
\begin{equation}
\label{4.12}
 \delta  m^2_{\lambda_{eff}}  \;  \simeq  \; 
\frac{\lambda_{eff}}{64 \pi^2} \,  \left [  \Lambda_h^2 - m_H^2 \;  ln \left(  \frac{\Lambda_h^2+m_H^2}{m^2_H } \right ) \right ]  \; ,
\end{equation}
\begin{equation}
\label{4.13}
 \delta  m^2_t  \;  \simeq  \; 
\frac{\kappa \, \lambda_t^2}{32 \pi^2} \,  \left [ - \Lambda_h^2 + 3 \, m_t^2 \, ln \left(  \frac{\Lambda_h^2+m_t^2}{m_t^2 } \right ) 
 \; + \;  \frac{ 3 \, m_t^4 }{\Lambda_h^2+m_t^2} - 3 \, m_t^2 \right ] \;  . 
\end{equation}
To evaluate $\lambda_{eff}$ and $\delta m_h$ we set:
\begin{equation}
\label{4.14}
 \Lambda_h \; \simeq \; m_H \;  \simeq \; 730 \; {\text GeV} \; \;  \; , \; \; \; sin^2 \, \theta_W \; \simeq  \; 0.223 \; .
\end{equation}
Concerning the mass scale $\mu$, we assumed $ 1 \; {\text GeV} \; \lesssim \; \mu \; \lesssim \; 100 \; {\text GeV}$
and obtained:
\begin{equation}
\label{4.15}
m_h \; \simeq \;  50 \; - \; 60  \; {\text GeV} \; .
\end{equation}
We are led, thus, to the remarkable prediction that there are two kind of elementary excitations of the Higgs condensate that
resemble closely a heavy Higgs boson with mass around 750 GeV and a light Higgs boson with mass of order 100 GeV.
Obviously, the light Higgs boson is naturally identified with the new narrow resonance at 125 GeV. Note, however,
that according to Eq.~(\ref{4.15}) we have:
\begin{equation}
\label{4.16}
m_h^{exp} \; \simeq \; 2.5 \;  m_h \; .
\end{equation}
We believe that the difference between the theoretical estimate Eq.~(\ref{4.15}) and the observed mass is due to the fact that our approximations
completely neglect quantum correlation effects. Indeed, in condensed matter it is well known that correlations lead to elementary quasi-particle
with an effective mass different from the "free" mass. Equation (\ref{4.16}) suggests that the scalar condensate behaves as a quantum liquid with
non-negligible correlations. These correlations are expected to affect appreciably the long-range phonon-like excitations. On the other hand,
we do not expect sizeable correlation effects on the roton-like excitations since these arise from localised disturbance of the
scalare condensate. \\
We would like to end this Section by attempting at least a qualitative estimate of the size of the correlation effects. We push further the analogy
with liquid helium by assuming that the role of the average distance between helium atoms is naturally played  by $d \sim 1/m_H$. 
We have seen that  the interactions of the roton-like scalar condensate  excitations with mass $m_H$ with the vector bosons and fermions
induce an effective positive quartic self-coupling. This repulsive short-range interaction will distort the condensate over a 
distance $D \sim  \frac{1}{\sqrt{\lambda_{eff}} \, v}$. The Bogoliubov's dilute gas approximation corresponds here to $D \gg d$
or, equivalently, $\lambda_{eff} \ll 1$. Indeed, one can easily check that $D \gtrsim 10 \, d$.  The distortion of the condensate by
quantum fluctuations will, in  turn, increase the inertia of the long-range phonon-like excitations. In fact, observing that
$<v| | \nabla \hat{h} |^2 |v> \; \sim \; \frac{h^2}{D^2}$, we see that the mass of long-range condensate excitations increases by 
$\sim 1/D \sim  \sqrt{\lambda_{eff}} \, v$. This should push the mass of the light Higgs boson closer to the experimental value.
\section{Phenomenology of the two Higgs bosons}
\label{s-5}
We have seen that the perturbations of the scalar condensate due to the quantum Higgs field behave as two independent massive
scalar fields in the dilute gas approximation that is the relevant regime for the LHC physics. To see what are the experimental signatures
of our proposal it is necessary to examine the interactions of the condensate elementary excitations. The most evident consequence of our
approach is the prevision of two Higgs bosons. The light Higgs boson is a natural candidate for the new LHC scalar resonance at 125 GeV.
Therefore we shall indicate our light Higgs boson with h(125). On the other hand, our previous phenomenological analysis of the
preliminary LHC Run 2 data in the golden channel~\cite{Cea:2019} suggested the presence of a broad scalar resonance with
central mass at 730 GeV. Accordingly, we shall denote the heavy Higgs boson with H(730). Note that this mass value is consistent
with the lattice determination Eq.~(\ref{3.10}). Obviously, these two Higgs bosons will interact with the gauge vector bosons.
We already pointed out~\cite{Cea:2017,Cea:2019} that the couplings of the Higgs condensate elementary excitations to the
gauge vector bosons are fixed by the gauge symmetries. As a consequence, both the Higgs bosons h(125) and H(730) will be
coupled to gauge bosons as in the usual perturbative approximation of the Standard Model. As concern the coupling
to fermion fields, if we admit the presence of the Yukawa terms in the Lagrangian, then, after taking into account Eq.~(\ref{4.1}),
we get:
\begin{equation}
\label{5.1}
{\hat{\cal L}}(x) \; = \;  \frac{\lambda_f}{\sqrt{2}} \; v \; \hat{\bar{\psi}}_f(x)    \hat{\psi}_f(x)   \;  + \;   
 \frac{\lambda_f}{\sqrt{2}}  \; \hat{\bar{\psi}}_f(x)    \hat{\psi}_f(x) \, \hat{H}(x)  \; ,
\end{equation}
where  $\hat{\psi}_f(x)$ is a generic fermion quantum field.  The first term in Eq.~(\ref{5.1}) gives the interaction of the massless fermion field
with the condensate, while the second term is the interaction of the quantum Higgs field with fermions. As is well known, the repeated scatterings
of the massless fermions with the (almost) uniform Higgs condensate generate a fermion mass given by:
\begin{equation}
\label{5.2}
m_f \; = \;  \frac{\lambda_f}{\sqrt{2}} \; v \; .
\end{equation}
In perturbation theory   $\hat{H}$(x) is an elementary quantum field. So that the coupling of the elementary Higgs field to fermions is related
to the fermion mass by:
\begin{equation}
\label{5.3}
\lambda_f \; = \;  \frac{\sqrt{2} \, m_f }{v} \; .
\end{equation}
However, in our approach the scalar quantum field   $\hat{H}$(x) can create two different quasiparticles. In the dilute gas approximation
these quasiparticles can be described by two weakly-interacting elementary quantum fields, $\hat{h}$(125)(x)  and $\hat{H}$(730)(x),
except that particle creation and destruction operators must be replaced by the quasiparticle creation and destruction operators.
Therefore, instead of Eq.~(\ref{5.1}) we have:
\begin{equation}
\label{5.4}
{\hat{\cal L}}(x) = \frac{\lambda_f}{\sqrt{2}} v  \hat{\bar{\psi}}_f(x)    \hat{\psi}_f(x)    +    
\sqrt{Z^h_{wf}}  \frac{\lambda_f}{\sqrt{2}}  \hat{\bar{\psi}}_f(x)    \hat{\psi}_f(x)  \hat{h}(125)(x)   +  
\sqrt{Z^H_{wf}}  \frac{\lambda_f}{\sqrt{2}}  \hat{\bar{\psi}}_f(x)    \hat{\psi}_f(x)  \hat{H}(730)(x)  
\end{equation}
where  $Z^h_{wf}$ and   $Z^H_{wf}$ are wavefunction renormalisation constant~\cite{Nozieres:1962,Luttinger:1962} that, roughly,  take care
of the eventual mismatch in the overlap between the fermion and quasiparticle wavefunctions. Note that the gauge symmetries assure that
there are not renormalisations in the coupling of the quasiparticles to the gauge fields. This corresponds in condensed matter
to the well-known fact that a quasielectron has exactly the same electric charge of a free electron.  \\
A direct calculation of the wavefunction renormalisation constants is not easy. Nevertheless, we can fix these constants from a 
comparison with the experimental observations. After the end of the Run 2 at the Large Hadron Collider it resulted that the narrow
scalar resonance at 125 GeV were consistent with the perturbative Higgs boson of the Standard Model~\cite{Aad:2019,CMS:2019}.
In particular, the Yukawa couplings of the resonance at 125 GeV with the top and bottom quarks and with the $\tau$ lepton are
consistent with the theoretical predictions from perturbation theory. As a consequence, we are led to assume that:
\begin{equation}
\label{5.5}
Z^h_{wf}  \; \simeq \; 1 \; . 
\end{equation}
A remarkable consequence of Eq.~(\ref{5.5}) is that our light Higgs boson h(125) is indistinguishable from the perturbative Higgs boson. The unique
difference derives from the Higgs self-coupling.  In the perturbative approach the self-coupling is a free parameter related to the Higgs boson mass
by Eq.~(\ref{3.5}):
\begin{equation}
\label{5.6}
\lambda_{SM}  \; = \; \frac{3 \, m_h^2}{v^2} \; . 
\end{equation}
On the contrary, as discussed in Sect.~\ref{s-4}, in our approach the Higgs self-coupling can be estimate. In fact, we found in the lowest-order
approximation that $\lambda_{eff} \ll 1$. More precisely,  we have:
\begin{equation}
\label{5.7}
\frac{\lambda_{eff}}{\lambda_{SM}} \;  \lesssim\; 0.1 \; . 
\end{equation}
In principle Eq.~(\ref{5.7}) can be contrasted with the experimental observations. Indeed, the Higgs self-coupling gives rise to triple and quartic Higgs 
vertices with well-defined experimental signatures. The test of the quartic Higgs vertex probably is not possible even at the high luminosity
LHC. However, the triple Higgs vertex can be constrained experimentally from searches for double Higgs boson production. In fact, recently
the ATLAS Collaboration~\cite{ATLAS:2019} was able to set limits on the Higgs boson self-coupling by combining the single
Higgs boson analyses with the double Higgs boson analyses in several different decay channels using data at $\sqrt{s}$ = 13 TeV with
an integrated luminosity up to 79.8 fb$^{-1}$ for the single Higgs boson and up to 36.1 fb$^{-1}$ for the double Higgs boson.
By assuming that new physics affects only the triple self-coupling $\lambda_{HHH}$, they reported:
\begin{equation}
\label{5.8}
 - 2.3 \; <  \; \frac{\lambda_{HHH}}{\lambda_{SM}} \;  < \; 10.3  \;  \; \; \; \; \; \; \text{ATLAS}
\end{equation}
at the 95 \% confidence level. Likewise, in Ref.~\cite{CMS:2019} the CMS Collaboration set limits on the Higgs self-coupling by
combining measurements of the production and decay rates of the Higgs boson using the data set recorded at  $\sqrt{s}$ = 13 TeV
corresponding to an integrated luminosity of up to  137 fb$^{-1}$, depending on the decay channel:
\begin{equation}
\label{5.9}
 - 3.5 \; <  \; \frac{\lambda_{HHH}}{\lambda_{SM}} \;  < \; 14.5  \;  \; \; \; \; \; \; \text{CMS}
\end{equation}
at the 95 \% confidence level.  It is evident that to distinguish the perturbative Higgs boson from our proposal it is necessary
to increase considerably the integrated luminosity. 
\begin{figure}
\vspace{-0.0cm}
\begin{center}
\includegraphics[width=0.7\textwidth,clip]{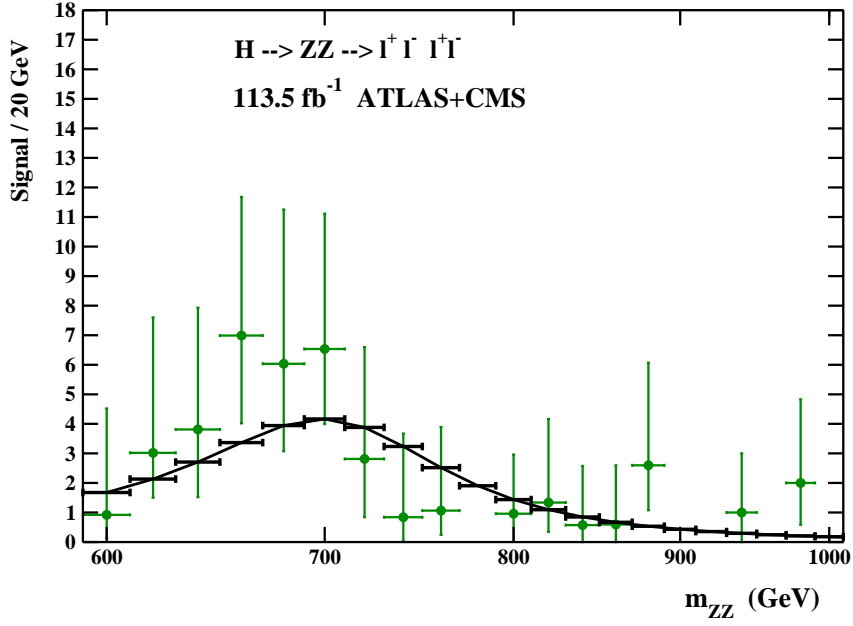}
\end{center}
\includegraphics[width=0.5\textwidth,clip]{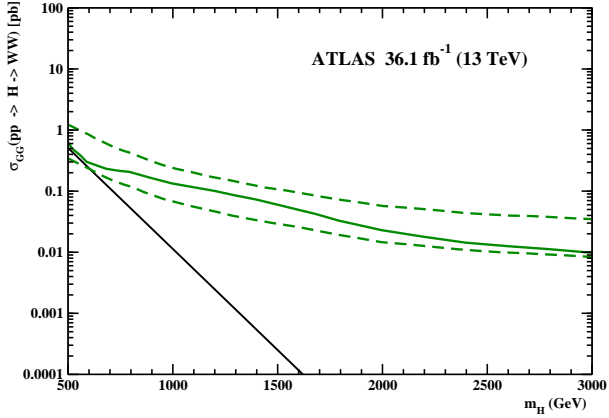}
\hspace{0.2 cm}
\includegraphics[width=0.5\textwidth,clip]{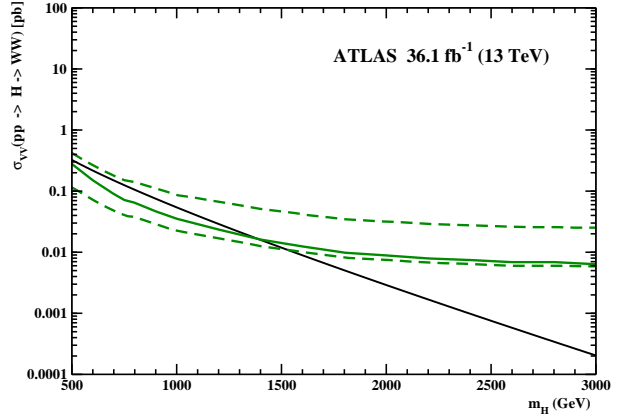}
\caption{\label{Fig3} (upper panel)
Comparison to the LHC data, (green) full points, of the distribution of the invariant mass $m_{Z Z}$ for the  process
 $H \; \rightarrow ZZ \; \rightarrow \ell \ell \ell \ell$   ($\ell = e, \mu$)  in the high-mass region   $m_{Z Z}  \gtrsim 600 \,  GeV$
with   the  expected signal histograms obtained assuming $\kappa \simeq \frac{m_h}{m_H}$. The data have been taken from
 Ref.~\cite{Cea:2019}. 
(lower panels) Limits on the production cross section times the branching fraction for the processes $pp \rightarrow H \rightarrow WW$. 
The data have been taken from Fig.~5 of Ref.~\cite{ATLAS:2019}. The dashed (green) lines demarcate  the 95 \% confidence level region
of the expected Standard Model background. The thick continuous (green) lines are the observed signal. The thin continuous black lines
are our theoretical estimate for the gluon-gluon fusion (left panel) and vector-boson fusion (right panel) production cross section times the
branching ratio Br(H $\rightarrow$ WW).
 }
\end{figure}
Fortunately,  our proposal can be more easily contrasted to observations by looking at the heavy Higgs boson H(730). Again, the
phenomenological signatures of the heavy Higgs boson H(730) depend on the couplings to gauge bosons and fermions. Since
the couplings to the gauge vector bosons are fixed by the gauge symmetries, it follows that the main decay modes of the 
heavy Higgs boson H(730) are given by the decays into W$^+$W$^-$ and Z$^0$Z$^0$ with~\cite{Cea:2019}:
\begin{equation}
\label{5.10}
Br(H(730) \rightarrow W^+ W^-) \; \simeq \; 2 \; Br(H(730) \rightarrow Z^0 Z^0) \; .
\end{equation}
Moreover, for a heavy Higgs boson the relevant fermion coupling is the Yukawa coupling to the top quark. Actually, in our previous phenomenological
analysis on the heavy Higgs boson proposal~\cite{Cea:2017,Cea:2019} we suggested that the top Yukawa coupling could be strongly suppressed
according to Eq.~(\ref{4.4}). We were led to this suppression from the results of a search for heavy neutral resonances produced by gluon-gluon
fusion and decaying into two massive vector bosons reported by both ATLAS and CMS Collaborations using the preliminary LHC data at
$\sqrt{s}$ = 13 TeV. 
Interestingly enough, comparing Eq.~(\ref{4.4}) with Eq.~(\ref{5.4}) we infer that:
\begin{equation}
\label{5.11}
\kappa \; = \;    Z^H_{wf}  \; .
\end{equation}
On general grounds, it is known that $0 < Z^H_{wf}  \le 1$. Indeed, from the comparison with the experimental observations we have
concluded that   $ Z^h_{wf}  \simeq 1$ and   $ Z^H_{wf}  = \kappa \ll 1$. Even though we cannot evaluate the wavefunction renormalisation
constant, we would like to present some arguments that make plausible the small value of $ Z^H_{wf} $. Indeed, we said that the
wavefunction renormalisation is determined by the mismatch between the overlap of the fermion and the quasiparticle wavefunctions.
Now, the light and heavy Higgs bosons are collective excitations corresponding to disturbances of the scalare condensate over a region
with size of order $D \sim 1/m_h$ and $d \sim 1/m_H$ respectively. Therefore, we expect that the heavy quasiparticle will suffer a more severe
mismatch with respect to the light quasiparticle:
\begin{equation}
\label{5.12}
 \frac{Z^H_{wf}}{ Z^h_{wf}} \; \simeq \; \frac{d}{D} \; \simeq \; \frac{m_h}{m_H} \; .
\end{equation}
Since  $Z^h_{wf} \simeq 1$, we get:
\begin{equation}
\label{5.13}
 Z^H_{wf}  \; = \; \kappa \; \simeq  \;    \frac{m_h}{m_H} \; \simeq \; 0.17 \; . 
\end{equation}
It should be stressed that our estimate Eq.~(\ref{5.13}) is, at best, a phenomenological educated guess. Nevertheless, it is
reassuring to see that Eq.~(\ref{5.13}) is close to the phenomenological parameter $\kappa$ used in Ref.~\cite{Cea:2019}. \\
At this point  it is necessary to check if our theoretical proposal of a heavy Higgs boson in consistent with the available
LHC data. Firstly, we have redone the analysis presented in Ref.~\cite{Cea:2019}. In Fig.~\ref{Fig3}, top panel, we display
the (unofficial)  combination, presented in Ref.~\cite{Cea:2019},  of the ATLAS and CMS data  in the golden channel.
The data are compared with the theoretical distribution, obtained following Ref.~\cite{Cea:2019}, with the parameter $\kappa$
given by Eq.~(\ref{5.13}). Looking at Fig.~\ref{Fig3} we see that our theoretical estimate is still in reasonable agreement with the data.
On the other hand, according to Eq.~(\ref{5.10}), the most stringent constraints come from the experimental searches for a heavy Higgs boson
decaying into two W gauge bosons. In fact, both the ATLAS~\cite{Aaboud:2018}    and CMS~\cite{Sirunyan:2019} Collaborations
presented the results on searches for a neutral heavy scalar resonance decaying into a pair of W boson using data at $\sqrt{s}$ = 13 TeV
and corresponding to an integrated luminosity of 36.1 fb$^{-1}$ and 35.9 fb$^{-1}$, respectively. Since the resulting limits set by the two LHC 
Collaborations are compatible,  we merely  present  the comparison with the ATLAS data. In  Fig.~\ref{Fig3}, bottom panels,
we display the observed limits at 95 \% confidence level on the heavy Higgs boson production cross section times the branching fraction 
$Br(H \rightarrow WW)$ for the gluon-gluon fusion (left panel) and vector-boson fusion (right panel) production mechanisms in
the narrow width approximation as reported in Ref.~~\cite{Aaboud:2018}. It should be remarked that, in the search for a heavy neutral 
resonance decaying into a WW boson pair, no significant excess of events beyond the Standard Model expected background were
found in the explored mass range. This could lead to stringent constraints on our proposal. To this end, following Ref.~\cite{Cea:2019}, in 
 Fig.~\ref{Fig3} we report our estimate for the product of the gluon-gluon fusion production cross section and vector-boson fusion
 cross section times the branching ratio for the decay of the heavy Higgs boson into two W vector bosons.
For the gluon-gluon fusion production mechanism we see that, in the relevant mass range, our theoretical cross section lies  below the observed 
limits. This means that, at the moment, there is not enough sensitivity to detect the signal in this channel. On the other hand, the
theoretical vector-boson fusion cross section falls within the $\pm 2 \sigma$  ranges around the expected limit for the 
 Standard Model background only hypothesis.
 Even though the absence of a signal in this channel could seem problematic, our theoretical proposal is still viable.
In fact, the suppression of the top Yukawa coupling implies that the expected signal for the gluon-gluon fusion production mechanism is well
below the uncertainties  of the expected background. Moreover, in the vector-boson fusion production mechanism the integrated luminosity
is too low to safely disentangle the expected signal out of the background. Nevertheless, we expect that 
with the full LHC  Run 2 data set there should be a  signal at least for the vector-boson fusion production mechanism.
\section{Summary and conclusions}
\label{s-6}
In the present paper we proposed to picture the Higgs condensate of the Standard Model as a quantum liquid analogous to
the superfluid helium. Our approach allowed us to uncover the spectrum of the elementary excitations.
We found that there are two different kind of condensate excitations that are similar to phonon and rotons in helium II.
We found that, in the weak interaction approximation,  the Higgs condensate excitations behave as two Higgs bosons with mass
around 100 GeV and 750 GeV respectively. The light Higgs boson was identified the the LHC narrow resonance at 125 GeV.
The heavy Higgs boson found preliminary evidence in our previous phenomenological analysis in the golden
channel of the preliminary LHC Run 2 data from ATLAS and CMS Collaborations. We have critically contrasted our
theoretical proposal to the available LHC data. We concluded that up to now the experimental observations are not yet 
in contradiction with the scenario of two Higgs bosons. 
We are confident that the full data set of the LHC Run 2 will corroborate our theoretical proposal.

\end{document}